\documentstyle[12pt,epsf]{article}

\makeatletter

\def\vereq#1#2{\lower3pt\vbox{\baselineskip1.5pt \lineskip1.5pt
\ialign{$\m@th#1\hfill##\hfil$\crcr#2\crcr\sim\crcr}}}

\makeatother

\title{
\begin{flushright}\normalsize
\vspace{0cm}
Internal Report~~\\
DESY L-02-02~~~\\
November 2002
\vspace{0.cm}
\end{flushright}
\begin{center}
\centerline{\large\bf An Approach to Automatic Indexing of Scientific
Publications}
\vspace*{2mm}
\centerline{\large\bf in High Energy Physics for Database SPIRES HEP}  
\end{center}
}

\author{A.V.~Averin$^{a)}$, L.A.~Vassilevskaya$^{b)}$\\
       {\small\it a) NSI, B.Tulskaya 52, 113191 Moscow, Russia} \\
       {\small\it (E-mail address: ava@ibrae.ac.ru)} \\
       {\small\it b) Deutsches Elektronen-Synchrotron DESY, D-22603
       Hamburg, Germany} \\
       {\small\it (E-mail address: lyuba@mail.desy.de)}  }

\date{}

\begin{document}

\maketitle

\begin{abstract}
\noindent
%
We introduce an approach to automatic 
indexing of e-prints based on a
pattern-matching technique making extensive use of an
Associative Patterns Dictionary (APD), developed by us. Entries in
the APD consist of natural language phrases with the same semantic
interpretation as a set of keywords from a controlled vocabulary.
The method also allows to recognize within e-prints formulae
written in \TeX{} notations that might also appear as keywords.
We present an automatic indexing system, AUTEX, which we have applied
to keyword index e-prints in selected areas in high energy physics
(HEP) making use of the DESY-HEPI thesaurus as a controlled vocabulary.
\end {abstract}

\section{Introduction}
\label{Sec:introduction}

Indexing is considered classically as an evaluation of documents,
leading to its content-based representation, which could subsequently
be accessed using a descriptor-based information retrieval system.
These descriptors (keywords) are taken as a rule from a well-defined
and controlled vocabulary. One expects that this controlled vocabulary
is periodically updated to be {\it en vogue}, i.e., it would reflect new
fields, and new trends and innovations within the established fields.\\

\noindent
Intellectual indexing implies at least two essential
steps: the conceptual grasp of the content of a
document (content analysis), and transcribing the content in terms of
index descriptors (taken, for example, from a thesaurus). The central
point is that the intellectual indexing process requires 
 an understanding by a human indexer of the
``aboutness'' \cite{Chung} of a document being indexed.
Historically, this was the only method of indexing the literature,
as all publishing and related bibliographical work was done with
papers. This classic form of document analysis is still widely used,
and is particularly important in the circumstances where the texts
are available in a non-electronic form or for non-text documents
(pictures, photographs, films etc.). The down side of the
intellectual method is that it is expensive, subject-oriented, and hence 
localized, and time consuming. Moreover, human-based judgment errors can 
not always be avoided.\\

\noindent
 In the case of High Energy
Physics (HEP), which is the primary area of research we shall
concentrate on, keyword indexing has been done at the German  High
Energy Physics
Laboratory DESY (Deutsches Elektronen-Synchrotron)~\cite{DESY} 
in Hamburg, which has developed and maintained the High Energy 
Physics Index (HEPI) since 1963.
The DESY-HEPI keywords~\cite{HEPI} are included in all records
(published papers, preprints, conference proceedings and books) in the
HEP literature database called SPIRES (Stanford Public Information REtrieval System),
maintained at the high energy physics laboratory SLAC (Stanford Linear
Accelerator Center) in California~\cite{SPIRES}. This keyword
indexing has been done exclusively using the so-called
intellectual method in the Scientific Documentation Group at DESY,
involving a team which includes some members with HEP background.
The thesaurus used in HEPI is based on approximately 2500 terms, and
is occasionally updated. With the growing use of computer-based methods
and the rapid  increase in the  number of scientific publications, most of
which are prepared in electronic form, and a well-maintained central
repository of e-prints in HEP and related fields now based at Cornell
University~\cite{Ginsparg}, there is an  overriding need to develop 
methods of automatic
indexing. In fact, automatic indexing is in the long term {\it the only
realistic and economically viable method} of keyword indexing
scientific documents.\\

\noindent
However, despite all the technological
progress and, what concerns HEP and several related fields, the
availability of e-prints from the electronic  archives dating
back to the pioneering work of Ginsparg done at the Los~Alamos National
Laboratory in New Mexico in 1991~\cite{Ginsparg}, there is no
fully automatic indexing system in HEP available yet. In the
recent past, some progress in this direction has been made at
CERN (the European Laboratory for Particle Physics) in Geneva.
The first of these approaches called SOCRATES  was an attempt to
incorporate software for automatic indexing \cite{Dallman/Meur},
based on constructing a set of the longest noun-type phrases from
the abstracts of documents. This approach was, however, abandoned
at the end of 2000 due to the difficulties of integrating its
software into the database system of the CERN library. This was
followed by a second research and development (R\&D) project
called HEPindexer~\cite{Raez/Dallman}, in which the first step aimed 
at the generation of the DESY-HEPI keywords has been implemented 
following a statistical method. The results of HEPindexer,
which was trained on some 2400 high energy physics papers and tested
so far on some 1200 additional documents, is quite impressive in
that the results were close to 60\% in precision and recall
benchmarked against the HEPI-SPIRES method of keyword
indexing~\cite{SPIRES}. However, HEPindexer is not yet in use for 
professional indexing of the HEP documents.\\

\noindent
In this report we describe our approach to automatic indexing of HEP
e-prints based on {\it associative pattern} matching. Central to 
our approach is the {\it Associative Patterns Dictionary}, which we
have  developed and encoded.
Our software package, AUTEX (AUTomatic indEXer) has been tested so
far in a limited area of HEP, namely neutrino physics, astrophysics
and physics of the axion. This choice is  based on our expertise and
HEP background. The goal of AUTEX indexer is to take a document (an 
e-print) as input and represent its contents as a set of keywords 
from the DESY-HEPI thesaurus. 
Judging from the number of e-prints in these fields, our project
covers at present about 15\% of the e-prints in HEP. This is also
reflected in the number of the keywords presently encoded in the AUTEX
database (approximately 550) compared to approximately 2500 of the
DESY-HEPI thesaurus. We have tested our software package on about
300 e-prints in these fields and have compared the results of our
automatic indexing with the ones generated using the SPIRES-HEP literature
database. Our system can be accessed also by remote users and is 
already in a position to be used for indexing work in the stated 
areas of HEP.  We describe the 
salient features of our automatic indexer AUTEX, showing some 
representative results of the index reports on e-prints together 
with comparisons
with the SPIRES HEP generated intellectual indexing results. Templates
of the AUTEX screen at various intermediate steps 
of automatic indexing are also shown. In future, we plan to enlarge
the applications to cover the entire fields of HEP, astrophysics and 
cosmology, and to include additional features in the indexing 
reports such as the titles of the papers, authors names, and their 
institutional affiliations. Further details and AUTEX report 
results can be seen in Ref.~\cite{Vassilevskaya}. 

\section{AUTEX - an Automatic Indexer \\
of High Energy Physics Documents}
\label{Sec:system}
\vspace{5mm}

We start by describing the basic components of the AUTEX system.
These include the terminology, the {\it Associative Patterns 
Dictionary}, an introduction to the basic principle on which
AUTEX is based, and the user interface. 

\subsection{Terminology}
To begin, one should outline some definitions.\\

{\large\bf Keywords List} \\

\noindent
{\it Keywords List} is a subset of a controlled vocabulary.
In the case of high energy physics it is assumed to be covered by
DESY-HEPI thesaurus.  A {\it keyword} can be either a separate word: \\
\hspace*{5,0mm} {\bf astrophysics} \\
\hspace*{5,0mm} {\bf neutrino} \\
or a combination of some words: \\
\hspace*{5,0mm} {\bf dispersion relations} \\
\hspace*{5,0mm} {\bf interpretation of experiments}.\\

{\large\bf Keychain} \\

\noindent
{\it Keychain} is composed of keywords written in one line and joined
by a comma and a blank as a delimiter:
 \[
\underbrace{\overbrace{\bf astrophysics}^{keyword}}_{keychain}
\]
\[
\underbrace{
\overbrace{\bf neutrino}^{keyword}, \,
\overbrace{\bf magnetic \,\, moment}^{keyword}
}_{keychain}
\]
\[
\underbrace{
\overbrace{\bf plasmon}^{keyword}, \,
\overbrace{\bf longitudinal}^{keyword}, \,
\overbrace{\bf dispersion \, \, relations}^{keyword} \,
}_{keychain}
\]
\[
\dots
\]

\noindent
In principle the
system is able to construct a keychain of any length (any
number of keywords). The keychain of length one is just
a keyword.
At the moment, keychains of length no more than two are used
{\it de facto} within the Scientific Documentation Group at DESY. It
is keychains -- not keywords -- that are assigned to a document during
indexing process. The resulting set of keychains forms a document
index that is incorporated into the SPIRES-HEP database. \\

{\large\bf Associative Patterns Dictionary (APD)} \\

\noindent
In its simplest	form an {\bf associative pattern} is thought of as any
English phrase that might also include a formula written in \TeX{}
notations:\\
\hspace*{5,0mm}	{\tt astrophysics} \\
\hspace*{5,0mm}	{\tt neutron stars} \\
\hspace*{5,0mm}	{\tt search for galactic dark matter} \\
\hspace*{5,0mm}	{\tt \verb|$\nu \to \nu \gamma$|} \\
\hspace*{5,0mm}	{\tt axion decay \verb|$\a \to e^+ e^-$|.} \\

\noindent
Given an associative pattern, we assign to it a set of keychains
defining the {\it meaning} of this pattern. That is, one gets a
semantic interpretation of the pattern in terms of keychains from
the controlled vocabulary. Applying an alternation metasymbol {\bf $|$}
(vertical bar) makes it possible to consider multiple associative
patterns of the form:

\begin{center}
{\tt energy dissipation $|$ dissipation of energy $|$ energy is 
dissipated}.\\ 
\end{center}

\noindent
In most cases each
constituent phrase in an associative pattern is a noun phrase.
However, in principle,  {\it any} other phrase is allowed.
Here are some examples of how the APD entrance may look like: 

\begin{center}
\begin{tabular}{llr}
$\fbox{\parbox{3.5cm}{
{\tt leptogenesis} 
}}$ 
& $\Longrightarrow$ 
$\fbox{\parbox{4cm}{
{\bf lepton, production} 
}}$ \\

\vspace{1mm} & \\

$\fbox{\parbox{5.8cm}{
{\tt abelian horizontal charge} {\bf $|$} \\
{\tt horizontal abelian charge}
}}$ 
& $\Longrightarrow$ 
$\fbox{\parbox{4.5cm}{
{\bf horizontal symmetry} \\
{\bf charge, abelian}
}}$ \\

\vspace{1mm} & \\

$\fbox{\parbox{6cm}{
{\tt axion decay into electron- positron pair} {\bf $|$} 
{\tt axion decay $a \to e^+ e^-$} {\bf $|$} 
{\tt electron-positron pair production by axion} 
}}$ 
& $\Longrightarrow$ 
$\fbox{\parbox{6cm}{
{\bf axion, leptonic decay} \\
{\bf electron, pair production} \\
{\bf axion $\to$ positron electron} 
}}$ \\

\vspace{1mm} & \\
$\fbox{\parbox{6.5cm}{
{\tt neutrino pair production by a virtual photon}  {\bf $|$}
$\gamma_{virt} \to \nu \bar \nu$ 
}}$ 
& $\Longrightarrow$ 
$\fbox{\parbox{6,7cm}{
{\bf neutrino, pair production} \\
{\bf neutrino, photoproduction} \\
{\bf photon, off-shell} \\
{\bf photon $\to$ neutrino antineutrino} 
}}$ 
\end{tabular}
\end{center}

\noindent
Therefore, each entrance in the APD represents a relation, i.e., an
{\it association} between an {\it idea} and a set of index terms.

\subsection{How does AUTEX work?}

As already stated, the AUTEX indexer takes a document (an e-print) as
input and represents its content as a set of keywords from DESY-HEPI
thesaurus. Of course, this thesaurus is a default option currently and
as the applications increase, the DESY-HEPI can be replaced by another 
thesaurus.
Rather than thinking of a document as natural language input text
(which is additionally wrapped by \TeX{} commands), we treat
it as a stream of characters and apply the
pattern-matching technique. At the present stage, the AUTEX
system does not perform aspects of natural language
processing which traditionally include part-of-speech
tagging and syntactic analysis. Instead, each associative
pattern from the APD tries to match a document content.
If it succeeds, the corresponding set of keychains, associated
with this pattern, is added into a document index.
So, the AUTEX indexer does not distinguish whether a phrase
is a noun phrase or it is something else. It just matches
patterns. \\

\noindent
The indexer actually consists of two phases of operation: {\it
preprocessing} and {\it pattern matching}. The first
one in our case involves undertaking some {\it preprocessing} of the input
text. The system provides a set of built-in {\it pointers} that
enable us to specify which parts of the article to work out.
Currently, the following six pointers are available: ``title'',
``abstract'', ``caption'', ``section'', ``conclusions'', and
``full-text''.
Once a pointer is set, the system parser
extracts the corresponding part of the document and includes it
into the text to be processed. These parts of the text usually
(though not always) contain an important information on the
purpose of investigations and/or the results obtained in the
paper. So, reading these parts of the article is expected to lead to
information on the text relevant for keyword indexing.
Of course, multiple choice is
possible and is used in the actual indexing work presented here.
Note also that the ``section'' pointer
instructs the parser to fetch titles of the various sections (not the
section contents themselves) including the titles of ``subsections'',
``subsubsections'', etc. as well. \\

\noindent
Another goal of the preprocessing is to prune irrelevant
\TeX{} commands from each part to be pointed out. For example,
the pattern {\tt strong magnetic field} has to match the occurrence
{\tt \verb|{\it strong}| magnetic field} as well. \\

\subsection{User Interface}

\noindent
For the AUTEX indexer to work it has to be supplied with an
appropriate look up dictionary -- APD in our case. It is obvious, that
the APD is of crucial importance for an effective and correct 
automatic indexing. In particular, how the  APD is populated and 
maintained will play a central role in indexing.
At the current stage
of development of the AUTEX project, APD is populated
``by hand''. This work is done by a subject specialist
making use of a graphic user interface provided by the system.
This interface enables navigation across the system and
facilitates the management of Keyword List, Keychain
List, APD, and the indexer itself. For example, within APD Manager,
the APD records can be accessed using a flexible filter.
The filter options include keying of a truncated string
of characters, alphabetical selection, and keychain
selection. This enables one to select all patterns associated
with a given set of keychains. Similar filters are
provided in Keyword Manager and Keychain Manager as well.

\section{Pattern Matching}

\noindent
The patterns used in AUTEX pattern matching are, in fact,
regular expressions in their simplest form. Let us suppose that we are 
creating a pattern that will succeed if an input text is about massless 
neutrino. It may be seen, for example,
that the pattern {\tt massless~neutrino} will not match the
occurrence {\tt massless~chiral~neutrinos}. Such cases are easily
resolved if one introduces the following regular expression 

\begin{center}
{\tt \verb|massless[ \w]+neutrinos?|} \\
\end{center}

\noindent
where the metacharacter {\sf [\,\,]} and the
quantifiers {\sf ?} and {\sf +} have their standard 
meaning in the extended regular expression syntax that is supported by 
{\sf egrep} routine.
The backslashed metacharacter {\tt \verb|\w|} matches a single
alphanumeric character.
This pattern will succeed whatever words occur between the words
{\tt massless} and {\tt neutrino}, but it will still fail on the
phrase, for example, {\tt the neutrino is assumed to be massless}.
To match this phrase too, we modify our pattern to yield

\begin{center}
{\tt \verb|massless[ \w]+neutrinos?|{\bf $|$}\verb|neutrinos?[
\w]+massless|} \\
\end{center}

\noindent
This improves pattern matching. However, it will occasionally also process
English text in which a certain pattern is matched though the sense in
which these words are used is different than what a certain {\it 
keychain} implies. For example, 
one may encounter the following text in a document:
{\tt neutrino decays into some particles one of which
has to be massless}.
The last pattern given above will match this text. However,
this will be a wrong matching in the sense of the content, as the text 
does not discuss  massless neutrino but rather a massless particle which 
is the decay product of a neutrino. \\

\noindent
Therefore, if we want that our pattern does not miss any relevant
sentence about massless neutrino and all matches correctly depict the
content of a document, then one has to add further discriminatory
characteristics in the pattern matching engine.
In particular, this engine has to be able to first understand the
grammar of the sentence. Regular expressions are extremely powerful
tool.
However, they are not sufficient for grammar understanding. So, we must
integrate into the system lexical and syntactic analyzers of
English.  The idea is to identify in the original text some
constituents so that they will be recognized by a limited set of
associative patterns, filtering out peripheral text portions. 
There is an additional complication in our case due to the \TeX{} 
formulae. This requires further development work and 
we hope to return to this problem in a future publication.
\\

\noindent
Note that we stay within the framework based on an human-made
APD rather than on any statistical model. That is, each associative
pattern is built by a human -- a subject specialist -- and each
association is assigned by an expert. This approach results in rather
accurate and reliable index. Our first experiments show that, at
this stage, only a few index terms are lost.
The system can be sufficiently trained eventually to reduce this loss to
a bare minimum. We have striven at reducing the appearance of
redundant or irrelevant terms in the index, and our studies show that
in this we have largely succeeded.\\

\section{Results and Concluding Remarks}

\noindent
We conclude in this section by  showing some representative results
from the AUTEX indexer. The results of keyword indexing ten 
documents from the arXiv.org e-print archive are shown in Appendix 
A. The indexing reports from the AUTEX indexer are compared with the
already existing SPIRES HEP reports, both of which use the DESY-HEPI
thesaurus. The papers discuss mostly neutrino physics, and in one case
physics of the axion.
Many more such comparisons can be seen in Ref.~\cite{Vassilevskaya}. These 
comparisons show that the AUTEX indexer does a comprehensive 
job in indexing e-prints, and some of the keywords in the AUTEX 
reports are missing from the SPIRES HEP generated reports. Certainly, 
there is some room for improvement in the AUTEX system as well, but 
the efficiency already reached in the areas covered by AUTEX is very 
high.\\

\noindent
We show some templates of the AUTEX screen at important stages 
of automatic indexing.
Figs.~1 and 2 show the examples
of the user interface screens corresponding to the two Managers
called {\it Keyword Manager} and {\it Keychain Manager}, 
respectively. 
The {\it Article Manager}, whose template from the AUTEX screen is shown 
in Fig.~3, enables one
to upload an article file, set up the 
pointers mentioned in section 2, submit selected articles in 
batch mode, and control the resulting index reports.
The AUTEX report includes the article index along with the 
information which indicates from which parts of the article
(title, abstract, etc.) each index term stems. A typical 
indexing result is shown in  Fig.~4. \\

\noindent
Within the AUTEX project, the high 
priority for us is to enlarge the
applications (done by enlarging the dictionary APD) to cover the
entire field of high energy physics, cosmology and astrophysics. 
This is a straight forward though time-consuming step. However, the 
results achieved so far are very encouraging and we trust that the goal of
developing a completely automatic system of indexing and retrieval of 
all e-prints in the fields of HEP, astrophysics and cosmology can be 
achieved. Already at its present stage AUTEX is a useful tool, and it can 
be used to help in indexing e-prints in some areas of HEP and 
astrophysics.\\
 
\noindent {\bf Acknowledgment}
L.V. would like to thank Dietmar Schmidt for his support and
encouragement during the course of this work. We thank David Dallman for
reading the manuscript. 

\newpage
\setcounter{figure}{0}
\begin{figure} 
\begin{center}
\epsfxsize=1.1\textwidth
\epsffile{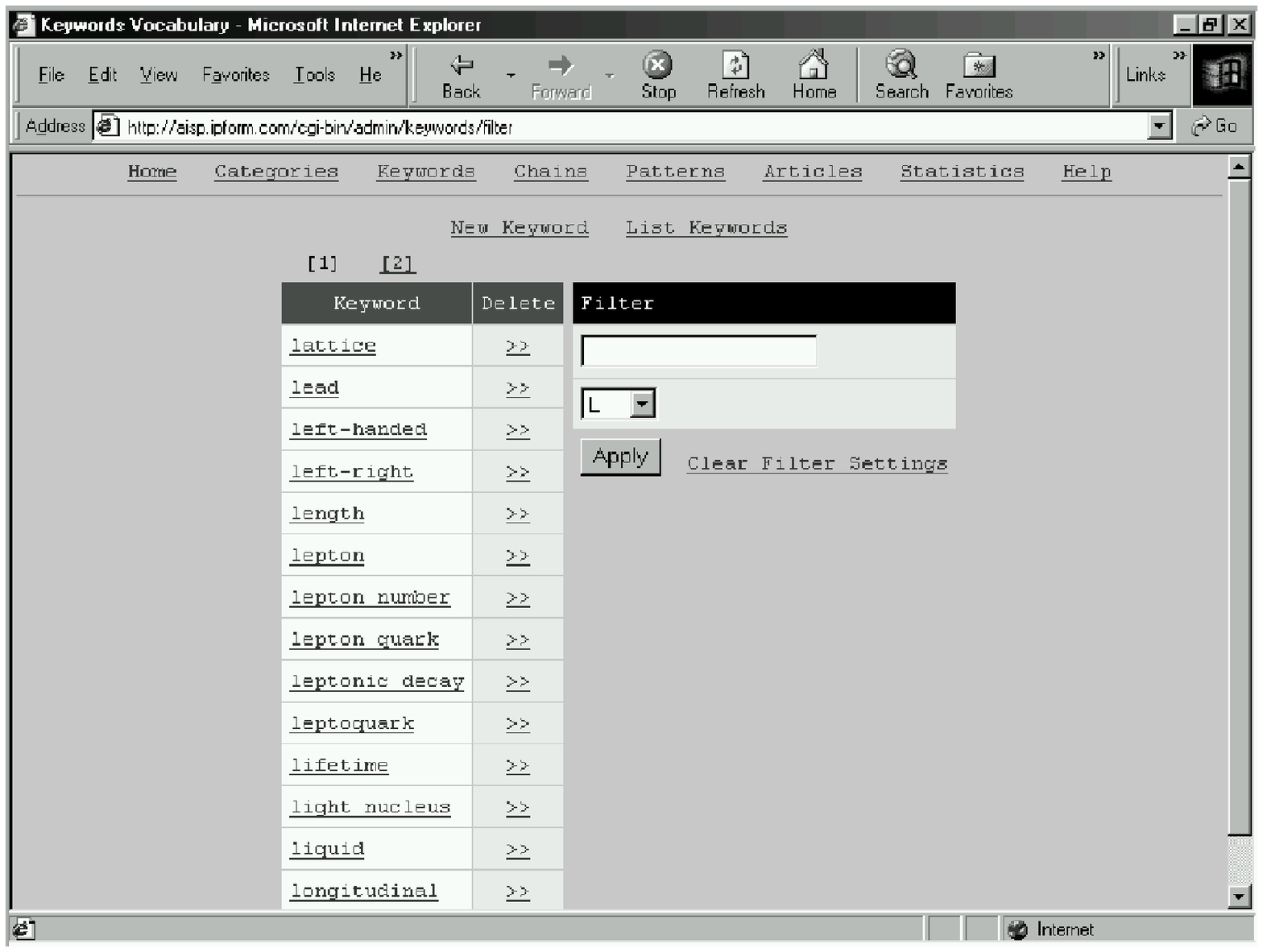}
\caption{ Base screen of the AUTEX Keyword Manager. Keywords are 
browsed in the most convenient way as a filtered list (shown on the 
left). The filter has two options: 1) one can choose a 
letter from the alphabet popup menu  for the system to select all 
keywords starting from this letter (here, the letter L), or 
2) one can specify a truncated string (more than one character). 
In this case, the system will return the list of keywords with these 
characters as leading. If both filter options are set, the list of 
keywords will satisfy both of them (logical AND operation). 
}
\end{center}
\label{Fig1}
\end{figure}

\setcounter{figure}{1}
\begin{figure} 
\begin{center}
\epsfxsize=1.1\textwidth
\epsffile{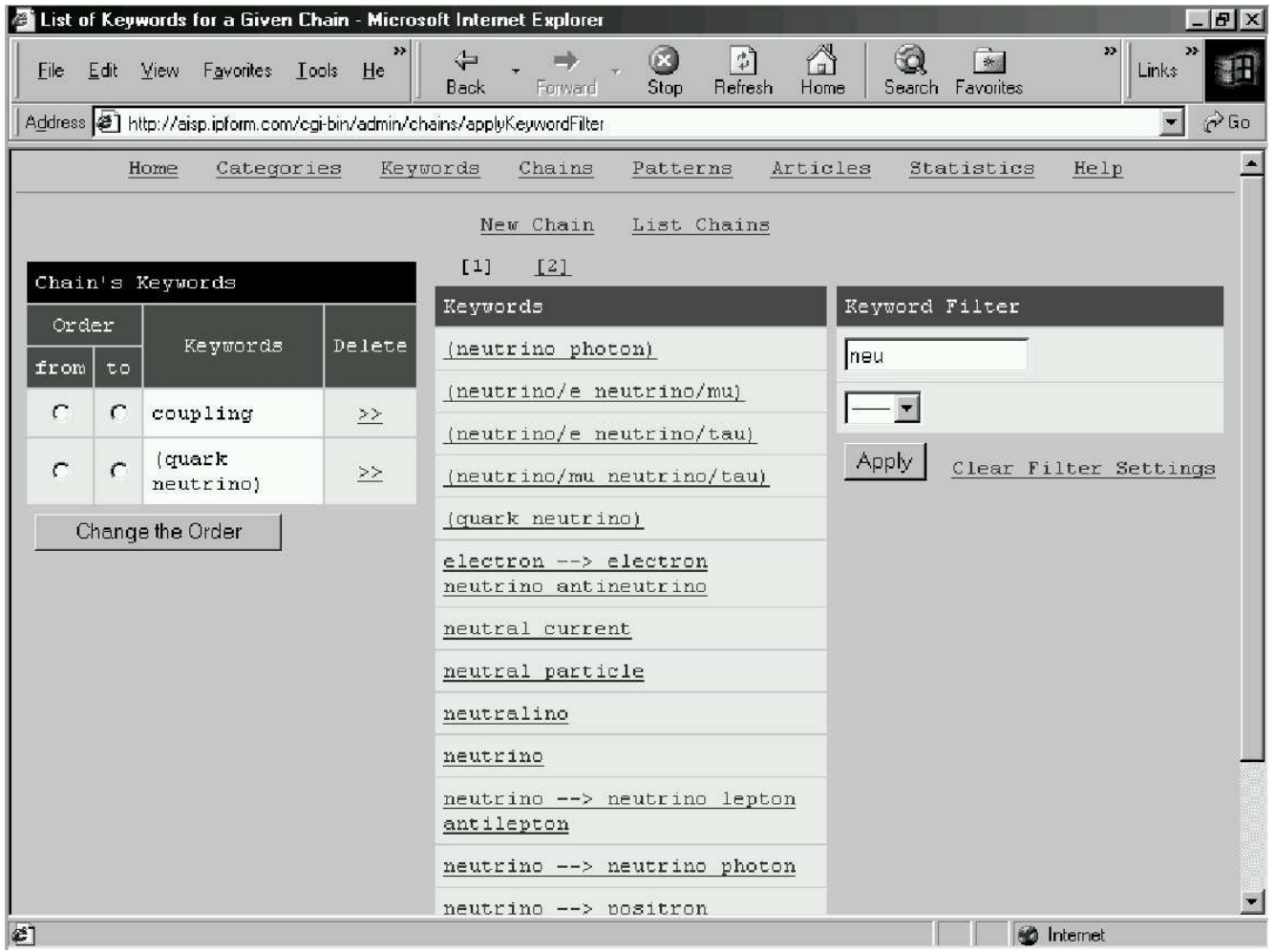}
\caption{A template from the AUTEX screen showing
%
a keychain creation/modification.
One can compose a keychain of any length  (any number of
keywords)  and change the order of keywords within the keychain. To
add a keyword  into a chain, one has to just press this keyword from
the filtered list. This keyword immediately appears in the chain.
}
\end{center}
\label{Fig2}
\end{figure}

\setcounter{figure}{2}
\begin{figure} 
\begin{center}
\epsfxsize=1.1\textwidth
\epsffile{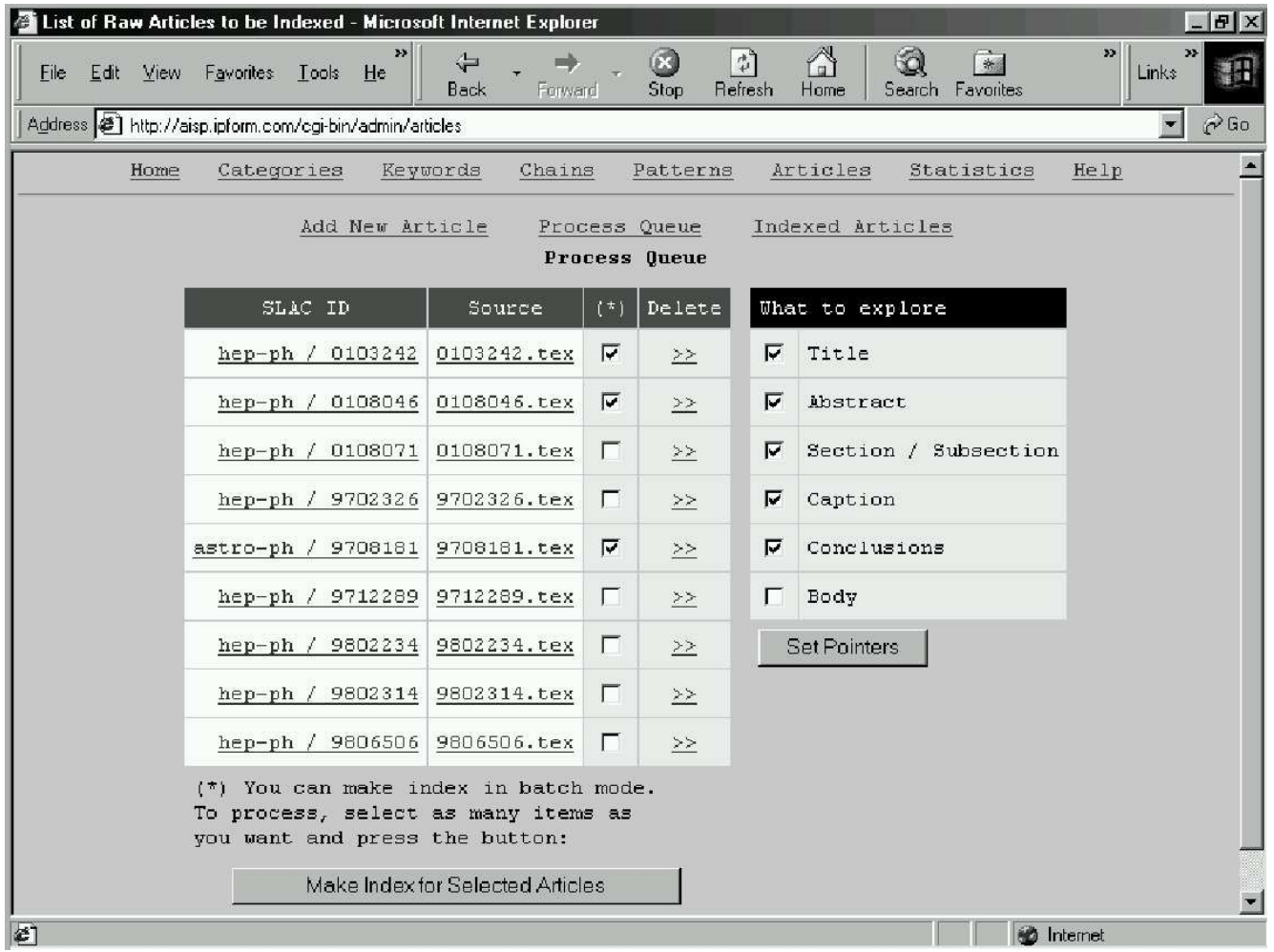}
\caption{A template from the AUTEX screen, showing a typical setup
with multiple pointers
%
along with the process queue, that is a list of references to the
e-prints that have already been uploaded into the system, but not
indexed yet. The e-prints from the process queue then can be indexed 
in the batch submission mode.
}
\end{center}
\label{Fig3}
\end{figure}

\setcounter{figure}{3}
\begin{figure} 
\begin{center}
\epsfxsize=1.1\textwidth
\epsffile{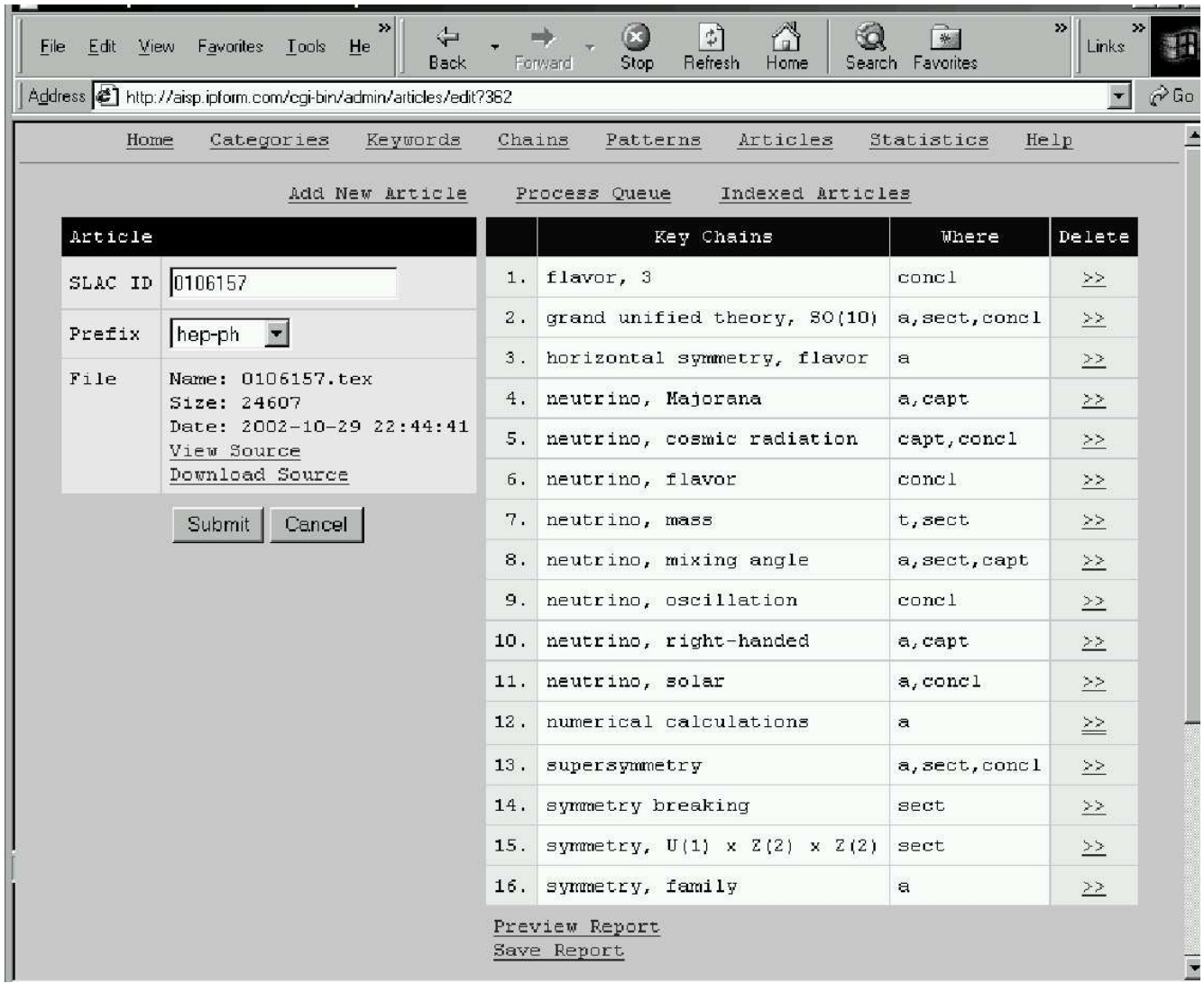}
\caption{A template from the AUTEX screen, showing the results of 
the indexed e-print (here hep-ph/0106157). 
The results can be 
previewed, corrected, printed, and stored in a file. One can edit an 
article profile that includes, at present, the SLAC ID and prefix.
In future, article title, authors names and addresses are planed to 
be incorporated into the profile.} 
\end{center}
\label{Fig4}
\end{figure}

\newpage
\appendix
\section{Appendix: AUTEX Reports (Examples)} 
\label{app:Glossary}

In this Appendix, we present some representative reports on 
subject keyword indexing generated by  
our automatic indexing system AUTEX and compare them with the 
already existing SPIRES reports. In all these examples, the papers refer 
to their E-Print archive numbers. The symbol ``0'' at the end of some 
reports means that this particular {\it keychain} has no relation to the 
paper. The {\it keychains} written below the horizontal line in some 
reports indicate mismatching {\it keychains} between the AUTEX and SPIRES
reports. 

\subsection{\, hep-ph/0106157}
\label{app:hep-0106157}

\begin{center}
{\small\bf Title:} 
{\bf GUT implications from neutrino mass} \\
\end{center}

\begin{abstract}
An overview is given of the experimental neutrino mixing results and types of 
neutrino models proposed, with special attention to the general features of 
various GUT models involving intra-family symmetries and horizontal flavor 
symmetries.  Many of the features are then illustrated by a specific 
$SO(10)$ SUSY GUT model formulated by S.M. Barr and the author which can 
explain all four types of solar neutrino mixing solutions by various choices 
of the right-handed Majorana mass matrix.  The quantitative nature of the 
model's large mixing angle solution is used to compare the reaches of a 
neutrino super beam and a neutrino factory for determining the small 
$U_{e3}$ mixing matrix element. \\
\end{abstract}

\begin{tabular}{lll}
\noindent
\underline{{\rm AUTEX Report:}} &  & \underline{{\rm SPIRES Report:}} \\ \\
 {\bf neutrino, mass  } & & {\tt neutrino, mass  } \\
 {\bf grand unified theory, SO(10) } & & 
 {\tt grand unified theory, SO(10) } \\
 {\bf symmetry, $U(1) \times Z_2 \times Z_2$ } & & 
         {\tt symmetry, $U(1) \times Z_2 \times Z_2$ } \\ 
 {\bf supersymmetry  } & & {\tt supersymmetry} \\
 {\bf horizontal symmetry, flavor } & & {\tt horizontal symmetry, flavor } \\
 {\bf symmetry, family } & & \\ 
 {\bf symmetry breaking  } & & {\tt symmetry breaking  } \\
 {\bf neutrino, mixing angle } & & {\tt neutrino, mixing angle } \\ 
 {\bf neutrino, right-handed } & & \\
 {\bf neutrino, Majorana } & & \\ 
 {\bf neutrino, flavor } & & \\
 {\bf flavor, 3  } & & \\
 {\bf neutrino, solar  } & & {\tt neutrino, solar  } \\
 {\bf neutrino, cosmic radiation  } & & \\
 {\bf neutrino, oscillation  } & & {\tt neutrino, oscillation  } \\
 {\bf numerical calculations  } & & {\tt numerical calculations  } \\
\end{tabular}

\newpage
\noindent
\subsection{\, hep-ph/0103176}
\label{app:hep-0103176}

\begin{center}
{\small\bf Title:} 
{\bf Neutrino kinetics in a magnetized dense plasma} \\ 
\end{center}

\vspace{-5mm}

\begin{abstract} 
The relativistic kinetic equations (RKE) for lepton 
plasma in the presence of a strong external magnetic field are
derived in Vlasov approximation. The new RKE for the electron spin
distribution function includes the weak interaction with neutrinos
originated by the axial vector current ($\sim c_A$) and provided by the
parity nonconservation. In a polarized electron gas Bloch equation
describing the evolution of the magnetization density perturbation is
derived from the electron spin RKE being modified in the presence of
neutrino fluxes. Such modified hydrodynamical equation allows to obtain
the new dispersion equation in a magnetized plasma from which {\it the
neutrino driven instability of spin waves} can be found. It is shown that
this instability is more efficient e.g. in a magnetized supernova than
the analogous one for Langmuir waves enhanced in an isotropic plasma. \\
\end{abstract} 

\begin{tabular}{lll}
\noindent
\underline{{\rm AUTEX Report:}} &  & \underline{{\rm SPIRES Report:}} \\ \\
{\bf kinetics, relativistic} & &  {\tt neutrino, kinematics} \\
{\bf lepton, plasma} & & {\tt lepton, plasma}  \\
{\bf plasma, magnetic} & &  \\ 
{\bf magnetic field, external field} & & 
                                 {\tt magnetic field, external field} \\
{\bf magnetic field, high} & & {\tt magnetic field, high} \\
{\bf Vlasov equation} & &  {\tt Vlasov equation} \\
{\bf electron, spin } & & {\tt electron, spin } \\
{\bf neutrino, interaction} & & {\tt neutrino electron, interaction} \\
{\bf current, axial-vector} & &  \\    
{\bf parity, violation} & &  \\   
{\bf electron, gas} & &  {\tt electron, gas} \\
{\bf electron, polarization} & & {\tt electron, polarization} \\
{\bf density, perturbation} & & {\tt density, perturbation}  \\
{\bf neutrino, beam} & & {\tt neutrino, beam} \\
{\bf neutrino, flux} & &  \\
{\bf dispersion relations} & &  {\tt dispersion} \\
{\bf stability} & &  {\tt stability} \\

{\bf astrophysics, supernova} & &  \\ 
{\bf lepton, current}  & & \\ 
{\bf current, conservation law}  & & \\ \hline
 & & {\tt neutrino, form factor} \\
 & & {\tt neutrino, charge} \\
 & & {\tt charge, induced} \\ \\
\end{tabular}

\newpage
\noindent
\subsection{\, hep-ph/0106085}
\label{app:hep-0106085}

\begin{center}
{\small\bf Title:} 
{\bf Models of neutrino masses and mixings} \\
\end{center}

\begin{abstract}
We briefly review models of neutrino masses and mixings. In view of 
the existing experimental ambiguities many possibilities are still
open. After an overview of the main alternative options we focus on 
the most constrained class of models based on three widely
split light neutrinos within SUSY Grand Unification. \\
\end{abstract}

\begin{tabular}{lll}
\noindent
\underline{{\rm AUTEX Report:}} &  & \underline{{\rm SPIRES Report:}} \\ \\
{\bf neutrino, mass} & & {\tt neutrino, mass} \\
{\bf neutrino, mixing angle} & &  \\
{\bf neutrino, flavor} & & {\tt neutrino, flavor}  \\
{\bf flavor, 3} & &  \\
{\bf flavor, 4} & &  \\ 
{\bf  mass, texture} & &  \\ 
{\bf hierarchy} & & {\tt hierarchy}  \\
{\bf lepton number, violation} & & {\tt flavor, violation}  \\
{\bf supersymmetry} & &  {\tt supersymmetry} \\
{\bf grand unified theory, SU(5)} & & {\tt grand unified theory, SU(5)} \\
{\bf grand unified theory, SO(10)} & & 
  {\tt grand unified theory, SO(10)}  \\ 
{\bf space-time, higher-dimensional} & &  
  {\tt space-time, higher-dimensional} \\
{\bf horizontal symmetry, U(1)} & & {\tt horizontal symmetry}  \\
{\bf charge, abelian} & &  \\
{\bf seesaw model} & &  \\   
{\bf neutrino, oscillation} & &  {\tt neutrino, oscillation} \\
{\bf neutrino, sterile} & &  {\tt neutrino, sterile} \\ 
{\bf neutrino, solar} & &  \\   
{\bf neutrino, right-handed } & &  \\  
{\bf neutrino, Dirac} & &  \\  
{\bf neutrino, Majorana } & &  \\  
{\bf MSW effect} & &  \\   
{\bf astrophysics, missing-mass } & &  \\  \\ \hline
      & &  {\tt neutrino, mass difference} \\
\end{tabular}

\newpage
\noindent
\subsection{\, hep-ph/0107217}
\label{app:hep-0107217}

\begin{center}
{\small\bf Title:} 
{\bf Photon damping caused by electron-positron pair production in a 
     strong magnetic field} \\
\end{center}

\begin{abstract}
 Damping of an electromagnetic wave in a strong magnetic field 
is analyzed in the kinematic region near the
threshold of electron-positron pair production. Damping of the 
electromagnetic field is shown to be noticeably
nonexponential in this region. The resulting 
width of the photon $\gamma \to e^+ e^-$ decay is 
considerably smaller than previously known results. \\
\end{abstract}

\begin{tabular}{lll}
\noindent
\underline{{\rm AUTEX Report:}} &  & \underline{{\rm SPIRES Report:}} \\ \\
 {\bf photon, absorption} & & \\
 {\bf photon, leptonic decay} & & \\ 
 {\bf electron, pair production } & & 
   {\tt electron \underline{positron}, pair production}  \\ 
 {\bf magnetic field, external field} & & 
   {\tt magnetic field, external field}\\
 {\bf magnetic field, high } & & {\tt magnetic field, high } \\
 {\bf photon, dispersion relations} & & \\ 
 {\bf field equations, solution} & & \\
 {\bf threshold, pair production} &  & {\tt threshold} \\ 
 {\bf kinematics } & & \\
 {\bf photon, polarization } & & \\
 {\bf vacuum polarization, ($\to$) } & & \\
 {\bf magnetic field } & & \\
 {\bf photon, width } & & \\
 {\bf photon $\to$ positron electron } & & \\ \\ \hline
       & & {\tt electromagnetic field} \\ 
       & & {\tt (0) photon, energy loss} \\ 
       & & {\tt numerical calculations} \\ 
\end{tabular}

\newpage
\noindent
\subsection{\, hep-ph/0112171}
\label{app:hep-0112171}

\begin{center}
{\small\bf Title:} 
{\bf Neutrino superbeam and factory tests of grand unified model
predictions for the large mixing angle and LOW solar neutrino solutions} \\
\end{center}

\begin{abstract}
Within the framework of an SO(10) GUT model that can accommodate both the 
LMA and LOW solar neutrino mixing solutions by appropriate choice of the
right-handed Majorana matrix elements, we present explicit predictions for the 
neutrino oscillation parameters $\Delta m^2_{21}$, $\sin^2 2\theta_{12}$, 
$\sin^2 2\theta_{23}$, $\sin^2 2\theta_{13}$, and $\delta_{CP}$.  Given the 
observed near maximality of the atmospheric mixing, the model favors the 
LMA solution and predicts that $\delta_{CP}$ is small.  The suitability of 
Neutrino Superbeams and Neutrino Factories for precision tests of the two 
model versions is discussed. \\
\end{abstract}

\begin{tabular}{lll}
\noindent
\underline{{\rm AUTEX Report:}} &  & \underline{{\rm SPIRES Report:}} \\ \\
 {\bf neutrino, mixing angle } & & {\tt neutrino, mixing angle } \\ 
 {\bf grand unified theory, SO(10)} & & {\tt grand unified theory, SO(10)} \\ 
 {\bf neutrino, solar} & & {\tt neutrino, solar} \\  
 {\bf neutrino, cosmic radiation} & & \\ 
 {\bf neutrino, right-handed} & & \\ 
 {\bf mass, Majorana} & & \\ 
 {\bf neutrino, oscillation} & & \\ 
 {\bf neutrino, mass difference} & & {\tt neutrino, mass difference} \\ 
 {\bf MSW effect} & & \\ 
 {\bf violation, CP} & & \\ 
 {\bf numerical calculations,} & & {\tt numerical calculations,} \\  
 {\bf interpretation of experiments} & & \\ \\ \hline
       & & {\tt neutrino, beam} \\     
       & & {\tt neutrino, particle source} \\     
\end{tabular}

\newpage
\noindent
\subsection{hep-ph/9910476}
\label{app:hep-9910476}

\begin{center}
{\small\bf Title:} 
{\bf Neutrino Oscillations in Electromagnetic Fields} \\
\end{center}

\begin{abstract}
Oscillations of neutrinos $\nu_L \leftrightarrow \nu_R$ in presence of 
an arbitrary electromagnetic field are considered. We introduce the
Hamiltonian for the neutrino spin evolution equation that accounts for
possible effects of interaction of neutrino magnetic $\mu$ and
electric $\epsilon$ dipole moments with the transversal
(in respect to the neutrino momentum) and also the longitudinal
components of electromagnetic field.  Using this Hamiltonian we
predict the new types of resonances in the neutrino oscillations 
$\nu_L \leftrightarrow \nu_R$ in the presence of the field of an 
electromagnetic wave and in combination of an electromagnetic wave
 and constant magnetic field. The possible influence of the 
longitudinal magnetic field on neutrino oscillations is emphasized. \\
\end{abstract}

\begin{tabular}{lll}
\noindent
\underline{{\rm AUTEX Report:}} &  & \underline{{\rm SPIRES Report:}} \\ \\
{\bf neutrino, oscillation} & & {\tt neutrino, oscillation} \\ 
{\bf neutrino, left-handed} & & \\ 
{\bf neutrino, right-handed} & & \\ 
{\bf electromagnetic field, external field } & & 
                      {\tt electromagnetic field} \\ 
{\bf electromagnetic field, longitudinal} & & \\ 
{\bf electromagnetic field, transversal} & & \\ 
{\bf magnetic field, longitudinal} & & \\ 
{\bf neutrino, momentum} & & \\ 
{\bf neutrino, spin} & & {\tt neutrino, spin} \\ 
{\bf effective Hamiltonian} & & \\  
{\bf neutrino, magnetic moment} & &  {\tt neutrino, magnetic moment} \\ 
{\bf neutrino, electric moment} & & \\ 
{\bf moment, dipole} & & \\ 
{\bf resonance} & & \\ 
\end{tabular}

\newpage
\noindent
\subsection{\, hep-ph/9812408}
\label{app:hep-9812408}

\begin{center}
{\small\bf Title:} 
{\bf Field-induced axion decay $a \to e^+ e^-$ via plasmon} \\
\end{center}

\begin{abstract}
The axion decay $a \to e^+ e^-$ via a plasmon is investigated 
in an external magnetic field. The results we have 
obtained demonstrate a strong catalyzing influence of the field
as the axion lifetime in the magnetic field of order $10^{15}$~G
and at temperature of order 10~MeV is reduced to $10^{4}$~sec. \\
\end{abstract}

\begin{tabular}{lll}
\noindent
\underline{{\rm AUTEX Report:}} &  & \underline{{\rm SPIRES Report:}} \\ \\
{\bf axion, leptonic decay } & & {\tt axion, leptonic decay} \\ 
{\bf electron, pair production} & & \\  
{\bf magnetic field, external field} & & 
                {\tt magnetic field, external field} \\  
{\bf yield, enhancement} & & \\   
{\bf plasmon, propagator} & & \\  
{\bf axion, dispersion relations}  & & {\tt dispersion relations} \\   
{\bf plasmon, dispersion relations}  & & \\ 
{\bf plasmon, longitudinal}  & & \\  
{\bf axion, lifetime} & & {\tt axion, lifetime} \\  
{\bf temperature } & & \\ 
{\bf numerical calculations} & & {\tt numerical calculations} \\  
{\bf Feynman graph}  & & \\  
{\bf axion $\to$ positron electron } & & \\ \\ \hline
{\sl coupling, (axion photon)} & & \\   
  & &  {\tt plasma}    

\end{tabular}

\vspace{7mm}

\noindent Note, that the keychain {\sl ``coupling, (axion photon)''} 
is not present in the {\rm AUTEX Report}, but it will be included 
in the next stage of the system development. 

\newpage
\noindent
\subsection{hep-ph/9710219}
\label{app:hep-9710219}

\begin{center}
{\small\bf Title:} 
{\bf Neutrino transitions $\nu \to \nu \gamma$,
       $\nu \to \nu e^+ e^-$ in a strong magnetic field 
       as a possible origin of cosmological $\gamma$-burst} \\
\end{center}

\begin{abstract}
The high energy neutrino transitions with the photon and
 electron-positron pair creation in a strong magnetic field  in the
 framework of the Standard Model are investigated. The process
 probabilities and the mean values of the neutrino energy and momentum
 loss are presented. The asymmetry of outgoing neutrinos, as a
 possible source of sufficient recoil ``kick'' velocity of a remnant
 and the emission of $e^+ e^-$-pairs and $\gamma$-quanta in a ``polar
 cap'' region of a remnant, as a possible origin of cosmological
 $\gamma$-burst are discussed. \\ 
\end{abstract}

\begin{tabular}{lll}
\noindent
\underline{{\rm AUTEX Report:}} &  & \underline{{\rm SPIRES Report:}} \\ \\
{\bf neutrino, transition} & & \\
{\bf neutrino, radiative decay} & & {\tt neutrino, radiative decay} \\ 
{\bf neutrino, leptonic decay} & & {\tt neutrino, leptonic decay} \\    
{\bf electron, pair production} & & {\tt lepton, pair production} \\
{\bf flavor, conservation law} & & \\                
{\bf neutrino, relativistic} & & \\
{\bf magnetic field, high } & & {\tt magnetic field, external field} \\  
{\bf electroweak interaction} & & {\tt electroweak interaction} \\
{\bf neutrino, energy loss} & & {\tt neutrino, energy loss} \\
{\bf neutrino, energy-momentum} & & \\
{\bf neutrino, emission} & & \\
{\bf neutrino, momentum} & & \\
{\bf momentum, asymmetry} & & \\
{\bf astrophysics, supernova} & & {\tt astrophysics, supernova} \\
{\bf n, matter} & & {\tt n, matter} \\
{\bf parity, violation} & & \\ 
{\bf pulsar, velocity} & & \\ 
{\bf gamma ray burst} & & \underline{{\tt photon, cosmic radiation}} \\
{\bf numerical calculations} & & \\   
{\bf neutrino $\to$ neutrino photon} & & 
               {\tt neutrino $\to$ neutrino photon} \\
{\bf neutrino $\to$ positron electron } & & 
               {\tt neutrino $\to$ positron electron } \\
\hspace*{24mm} {\bf neutrino} & & \hspace*{26mm}{\tt neutrino} \\ \\\hline
 & & {\tt (0) neutrino, massive} \\ 
\end{tabular}

\newpage
\noindent
\subsection{\, hep-ph/9404289}
\label{app:hep-9404289}

\begin{center}
{\small\bf Title:} 
{\bf The Radiative decay of a high-energy neutrino in the Coulomb
     field of a nucleus} \\
\end{center}

\begin{abstract}
In the framework of the Standard Model with lepton mixing the radiative
decay $\nu_i \rightarrow \nu_j \gamma$ of a neutrino of high ($E_\nu \sim
100 \, GeV$) and super-high ($E_\nu \ge 1 \, TeV$) energy is investigated
in the Coulomb field of a nucleus. Estimates of the decay probability and
``decay cross-section'' for neutrinos of these energies in the electric
field of a nucleus permit one to discuss the general possibility of carrying
out a neutrino experiment, subject to the condition of availability in the
future of a beam of neutrinos of these high energies. Such an experiment
could give unique information on mixing angles in the lepton sector of
the Standard Model which would be independent of the specific neutrino
masses if only the threshold factor ($1 - m_j^4 / m_i^4$) was not close
to zero.
\end{abstract}

\begin{tabular}{lll}
\noindent
\underline{{\rm AUTEX Report:}} &  & \underline{{\rm SPIRES Report:}} \\ \\
{\bf neutrino, radiative decay} & & {\tt neutrino, radiative decay}\\  
{\bf flavor, violation} & & \\  
{\bf neutrino, relativistic} & & \\  
{\bf neutrino, massive} & & \\  
{\bf external field, Coulomb} & & \\  
{\bf nucleus, electric field} & & {\tt nucleus, electric field} \\  
{\bf electroweak interaction} & & {\tt electroweak interaction} \\  
{\bf lepton, interference} & & {\tt lepton, interference} \\  
{\bf lepton, mixing angle} & & {\tt lepton, mixing angle} \\  
{\bf mass, threshold} & & \\  
{\bf neutrino, beam} & & \\  
{\bf cross section, decay} & & {\tt cross section, decay} \\  
{\bf neutrino  $\to$ neutrino photon} & & \\  
\end{tabular}

\newpage
\noindent
\subsection{\, astro-ph/9812366}
\label{app:astro-9812366}

\begin{center}
{\small\bf Title:} 
{\bf Neutrino emission due to Cooper pairing of nucleons in 
cooling neutron stars} \\
\end{center}

\begin{abstract}
The neutrino energy emission rate due to formation
of Cooper pairs of neutrons and protons in the superfluid
cores of neutron stars is studied. The cases of singlet-state pairing 
with isotropic superfluid gap and triplet-state
pairing with anisotropic gap are analysed.
The neutrino emission due to the singlet-state pairing of protons
is found to be greatly suppressed with respect to the
cases of singlet- or triplet-state pairings of neutrons.
The neutrino emission due to pairing of neutrons is shown to 
be very important in the superfluid neutron--star cores with the 
standard neutrino luminosity and with the luminosity
enhanced by the direct Urca process. It can greatly accelerate both, 
standard and enhanced, cooling of neutron stars with superfluid cores.
This enables one to interpret the data on surface temperatures of
six neutron stars, obtained by fitting the observed spectra with the
hydrogen atmosphere models, by the standard cooling with moderate 
nucleon superfluidity.
\end{abstract}

\begin{tabular}{lll}
\noindent
\underline{{\rm AUTEX Report:}} &  & \underline{{\rm SPIRES Report:}} \\ \\
{\bf neutrino, energy} & & \\
{\bf neutrino, emission} & & {\tt neutrino, emission} \\
{\bf astrophysics} & & {\tt astrophysics, model} \\
{\bf n, matter} & &  {\tt n, matter} \\
{\bf n, pair} & & \\
{\bf p, pair} & & \\
{\bf pair, Cooper} & & \\
{\bf nucleon, superfluid} & & {\tt superfluid} \\
{\bf superfluid, gap} & & \\
{\bf anisotropy} & & \\
{\bf neutrino, luminosity} & & {\tt neutrino, luminosity} \\
{\bf Urca process} & & \\
{\bf star, energy loss} & & \\
{\bf temperature, surface} & & \\
{\bf critical phenomena, temperature} & & \\
{\bf redshift} & & \\
{\bf temperature, dependence} & & {\tt temperature, dependence} \\
{\bf bremsstrahlung} & & \\
{\bf numerical calculations} & & 
  {\tt numerical calculations,} \\
 & &  {\tt interpretation of experiments} \\
\end{tabular}


\newpage
\addcontentsline{toc}{section}{References}
\begin{thebibliography}{100}
\label{Sec:bibliography}

\bibitem[Chung 1998]{Chung}
Yi-Ming Chung, William M. Pottenger, Bruce R. Schatz:\\
Automatic Subject Indexing Using an Associative Neural Network\\
(URL:http://www.cannis.uiuc.edu/chung98automatic.html).

\bibitem[Dallman/Meur 1999]{Dallman/Meur}
Dallman, D., Le Meur, J.Y.:
\newblock {Automatic Keywording of High Energy Physics.}
\newblock In 4th International Conference on Grey Literature: New
Frontiers in Grey Literature, Washington, DC, USA, 1999.

\bibitem[DESY]{DESY}
Deutsches~Elektronen-Synchrotron~DESY,~Hamburg \\
(URL: http://www.desy.de/).

\bibitem[Ginsparg 1991]{Ginsparg}
Ginsparg, P.:
\newblock {arXiv.org e-print archive (URL:
http://www.\-arXiv.\-org)}.
\newblock {The e-print archive arXiv.org was initially based at Los Alamos 
National Laboratory.
It has followed its creator, Paul Ginsparg, to Cornell University in
Ithaca, New York,  last year, but it is still often referred to as the 
LANL archive}.

\bibitem[HEPI]{HEPI}
The HIGH ENERGY PHYSICS INDEX
\newblock {Keywords 1996/1997}
\newblock Internal Report, DESY L-96-01, January 1996.

\bibitem[R{\'a}ez/Dallman 2001]{Raez/Dallman}
R\'aez, a.M., Dallman, D.:
\newblock {Experiences in Automatic Keywording of Particle Physics
Literature.}
\newblock High Energy Physics Libraries Webzine, Issue 5, 2001
(URL: http://library.cern.ch/HEPLW/5/papers/3/).

\bibitem[SPIRES]{SPIRES}
Stanford Linear Accelerator Center
\newblock SLAC Library, SLAC SPIRES HEP Datebase
(URL: http://www.slac.stanford.edu/spires/hep).

\bibitem[Vassilevskaya 2002]{Vassilevskaya}
Vassilevskaya, L.~A.; 
An Approach to Automatic Indexing of Scientific Publications in High 
Energy Physics for Database SPIRES-HEP;
DESY Report DESY L-02-01 , November 2002.


\end{thebibliography}
\end{document}